\documentstyle{elsart}

\def\beq{\begin{equation}}
\def\eeq{\end{equation}}

\def\bea{\begin{eqnarray}}
\def\eea{\end{eqnarray}}
\def\ba{\begin{array}}                  %array
\def\ea{\end{array}}

\begin{document}
\begin{frontmatter}
\title{ Space-Time Non-Commutativity near Horizon of a Black Hole}
\author{ M. Martinis and V. Mikuta-Martinis}
\address{Theoretical Physics Division,\\Rudjer Bo\v skovi\' c Institute, 10001 Zagreb, Croatia  \\
E-mail: martinis@irb.hr, vmikuta@irb.hr}

\begin{abstract}
We consider  dynamics of a quantum scalar field, minimally coupled
to classical gravity, in the near-horizon region of a
Schwarzschild black-hole. It is described by a static Klein-Gordon
operator which in the near-horizon region reduces to a scale
invariant Hamiltonian of the system. This Hamiltonian is not
essentially self-adjoint, but it admits a one-parameter family of
self-adjoint extension. The time-energy uncertainty relation,
which can be related to the thermal black-hole mass fluctuations,
requires explicit construction of a time operator near-horizon. We
present its derivation in terms of generators of the affine group.
Matrix elements involving the time operator should be evaluated in
the affine coherent state representation.

PACS: 03.65 Fd, 02.30.Tb, 04.70.-s, 04.62.+v, 04.60.-m

\end{abstract}
\end{frontmatter}
\newpage
\section{Introduction}
Einstein's general theory of relativity \cite{Nic} describes
gravity as a manifestation of the curvature of spacetime. A
fundamental instability against collapse implies the existence of
black holes as stable solutions of Einstein's equations. A black
hole is formed if a massive object (e.g. a star) collapses into an
infinitely dense state known as a singularity. In this picture the
curvature of spacetime becomes extreme and prevents any particle
even light from escaping to infinity. A black hole may have
several horizons that fully characterize its structure.

The simplest three-dimensional geometry for a black hole is a
sphere (known as a Schwarzschild sphere), its surface defines the
event horizon. In the case of a spherical black hole, with $
R_{\mu \nu} = 0 $, all horizons coincide at the Schwarzschild's
critical radius $r_s = 2GMc^{-2}$ .

The Quantum Field Theory (QFT) in curved spacetime with classical
event horizon is, however, troubled by the singularity at the
horizon \cite{tHooft}. This problem may be solved by treating the
black hole  as a quantum state which implies that the energy of
the black hole and its corresponding time do not commute at the
horizon \cite{Bai}.

In this picture we study the dynamics of a scalar field in the
near-horizon region described by a static Klein-Gordon (KG)
operator which in this case becomes  the Hamiltonian of the
system. The dynamics of a scalar field in the near-horizon region,
and its associated SO(2,1) conformal symmetry have been studied in
many papers
\cite{Cam},\cite{Stro},\cite{Gov},\cite{Gib},\cite{Birm} in which
a complete treatments of conformal quantum mechanics and of
near-horizon symmetry were made. In this letter, we present the
explicit construction of the time operator in the near-horizon
region in terms of the generators of the affine group, and discuss
its self-adjointness \cite{Mart}.

\section{Scalar Field in the Near-Horizon Region}

The Schwarzschild geometry of a static spherical black hole is
described by the metric ($c= \hbar = G = 1$)

\begin{equation}
 ds^2 = - f(r)dt^2 + [f(r)]^{-1}dr^2 + r^2d\Omega ^2 ,
\end{equation}

where $rf(r) = r - r_s$, and  $d\Omega ^2 = d\theta ^2 + sin^2
\theta d\varphi ^2$. Near horizon ($r \sim r_s$), $f(r)$ behaves
as $f(r)\sim 2\kappa (r-r_s)$, where $\kappa = 1/2r_s$ denotes the
surface gravity. The equation of motion of a free scalar field
$\Phi (x)$ in this background metric  is
\begin{equation}
  - \frac{1}{f} \Phi _{tt} + f \Phi _{rr} + (f ' + 2f/r)\Phi _{r} + \triangle _{\Omega}\Phi/r^2 - m^2\Phi
 =  0.
\end{equation}

 By separating the time and angular variables in $ \Phi (x) = e^{-it\omega}\phi
(r)Y_{lm}(\Omega )$, the mode $\phi $ with
 angular momentum $l$ and frequency $\omega $ satisfies the
 equation \cite{thooft2},\cite{Gup},\cite{Cam2}
 \beq
 \frac{1}{f}\omega ^2 \phi  + f \phi _{rr} + (f ' + 2f/r)\phi _{r}
 -
\frac{l(l+1)}{r^2}\phi - m^2\phi  =  0.
 \eeq

In the near-horizon region, the Schroedinger-like equation in the
 variable $ x = r - r_s $ can be studied. For very small values of $x$,
 the angular and mass terms can be neglected \cite{thooft3}. In terms of a new
 field variable $\psi = \sqrt{x}\phi$, the KG equation in the near-horizon  region, for small
$x$, reduces to  a scale  invariant   Schroedinger equation
\begin{equation}
[\frac{d^2}{dx^2} + \frac{(\frac{1}{4} + \Theta ^2)}{x^2}] \psi(x)
= 0
\end{equation}

where  $\Theta  = r_s\omega $. This reduction to a
Schroedinger-like equation can be  described by an effective
attractive  inverse square potential $ 2V(x) = -(1/4  + \Theta ^2)
x^{-2}$ which is conformally invariant with respect to the
near-horizon variable $x$. The corresponding quantum Hamiltonian
is
\begin{equation}
\hat{H} =  \frac{1}{2}(p^2 + \frac{g}{x^2} ),
\end{equation}
 where $ g = - (1/4  + \Theta ^2)$ is supercritical for all
 nonzero frequencies $\omega $ \cite{Cam3}. The static black hole
 Hamiltonian $H_{bh}$ near horizon can then be defined as
 \beq
 \hat{H}_{bh} = M + \hat{H}.
\eeq

\section{Time Operator and its Relation to Spacetime Non-commutativity}

The coordinate singularity associated with horizon may be overcome
by assuming spacetime non-commutativity on the horizon, the so
called Quantum Horizon \cite{tHooft}, \cite{Bai}. If a black hole
is considered as a quantum state, its energy $\hat{H_{bh}}$ and
its conjugate time $\hat{t}$ are expected to become conjugate
operators obeying formally
\begin{equation}
[\hat{H},\hat{t}] = i.
\end{equation}
Since $\hat{H}$ is x-coordinate dependent,  we expect spacetime
noncommutativity, $[\hat{t},x] \neq 0$. What is $\hat{t}$ as an
operator? Due to Pauli theorem \cite{Pauli}no such self-adjoint
operator should exist if the spectrum of the self-adjoint
Hamiltonian is semibounded or discrete. In quantum theory
$\hat{H}$ is  essentially self-adjoint only for $g > 3/4$ in  the
domain
\begin{equation}
{\cal D}_0 = \{ \psi \in L^2({\cal R}^{+},dx), \psi(0) = \psi'(0)
= 0 \}
\end{equation}
In this domain it has a continuous spectrum for $g \ge 3/4$ with
$E
> 0$ but no ground state at $E = 0$ \cite{Cam}.

For $g \leq 3/4$, the Hamiltonian is not essentially self-adjoint
\cite{Gov},\cite{Birm}, but it admits a one-parameter family of
self-adjoint extension labeled by a $U(1)$ parameter $e^{iz}$,
where $z$ is a real number, which labels the domains ${\cal D}_z$
of the extended Hamiltonian. The set ${\cal D}_z$ contains all the
vectors in ${\cal D}_0$, and vectors of the form $\psi_z =
\psi_{+} + e^{iz}\psi_{-}$.

For $g = - \frac{1}{4}$, as pointed out by Moretti and Pinamonti
\cite{Moretti}, the analysis of the spectrum of $H_z$  in some
papers \cite{Stro},\cite{Gov},\cite{Gib},\cite{Birm} is not
completely correct. By interpreting the logarithm, appearing in
$\psi_z $ near $x \sim 0$, as a one-valued function Moretti and
Pinamonti were able to show that for each $H_z$ with $z\neq 0$
there is exactly one proper bound state in ${\cal D}_z$:
 \beq
\psi_z (x) = N_z \sqrt{x} K_0\left( \sqrt{ E_z} x\right), \eeq
with the eigenvalue
 \beq
E_z = {\exp}\left[\frac{\pi}{2} {\cot} \frac{z}{2}\right],
 \eeq

 where  $N_z$ is a
normalization factor, and $K_{0}$ is the modified Bessel function.
The horizon, in this picture, is located at $x = 0$ where the wave
function $\psi _z$ vanishes. For $ x \sim 0 $ the logarithmic term
in (9) vanishes if $ x = x_z = 1/{\sqrt{E_{z}}} $ and $ \psi_{z} $
exhibits a scale behavior of the type $\psi _z \sim \sqrt{x}$. In
order to achieve that $x_z$ belongs to the near horizon region,
$z$ should be close to $z = 0+\epsilon , 2\pi + \epsilon , 4\pi +
\epsilon, ... $, where $\epsilon \sim 0$ \cite{Birm}. Then, in a
band-like region $\Delta _z = [x_z(1-\delta ), x_z(1+\delta )]$,
where $ \delta \sim 0 $, all the eigenfunctions of $\hat{H_z}$
exhibit a scaling behavior. However, the zero mode solution to (5)
is obtained  only for discrete values of $z_n = 2(2n+1)\pi $,
where $n = 0,1,2,...$.

The asymptotic scale invariance of the KG operator in the near
horizon region implies that the Hamiltonian $\hat{H}$, and the
scaling operator $\hat{D} = -(xp + px)/4$  obey the affine algebra
type commutation relation
\begin{equation}
  [\hat{H},\hat{D}] =i\hat{H}.
\end{equation}
This algebra can be easily  extended to the full $ SO (2,1) $
conformal algebra by adding to the set $(\hat{H},\hat{D})$ the
conformal generator $ \hat{K} = x^2/2 $. In this case the
$g$-dependent constant quadratic Casimir operator is obtained

\beq
C_2  =  \frac{1}{2}(\hat{K}\hat{H} + \hat{H}\hat{K}) -
\hat{D}^2  =  \frac{g}{4} - \frac{3}{16}.
\eeq

If $\hat{H}^{-1}$ exists, we can formally construct a time
operator $\hat{t}$
 \cite{Mart}
\begin{equation}
\hat{t} =  \frac{1}{2}( \hat{D}\hat{H}^{-1} +
\hat{H}^{-1}\hat{D}),
\end{equation}
which  obeys the required commutation relation, $[\hat{H},\hat{t}]
= i$. Although, both $\hat{H}$ and $\hat{D}$ separately can be
made self-adjoint operators in the domain $L^{2} ({\cal R}^{+},
dE) $ it is not true for a $\hat{t}$-operator which contains
$\hat{H}^{-1}$ \cite{Kla}. It is clear that  $\hat{t}$ is not a
self-adjoint operator  in the domain $ L^{2} ({\cal R}^{+}, dE) $
where
 \begin{eqnarray}
 \hat{H} & \longrightarrow & E \nonumber \\
 \hat{D}& \longrightarrow & -i(E \frac{d}{dE} + \frac{1}{2})
 \\
 \hat{t} & \longrightarrow & {-i\frac{d}{dE}}. \nonumber
 \end{eqnarray}
Following  Klauder \cite{Kla}, the commutation relation between
$\hat{H}$ and $\hat{t} $ should be considered on the Hilbert space
spanned by the affine coherent states.They are defined by
\begin{equation}
\mid \tau, \lambda \rangle = e^{ i \tau \hat{H}  }\, e^{ -2i (\ln
\lambda ) \hat{D} } \mid \eta \rangle
\end{equation}
where $\mid \eta \rangle$ is a fiducial vector chosen in such a
way as to satisfy $ \langle \eta \mid \hat{H}^{-1}\mid \eta
\rangle < \infty $. For example, if we choose $ \eta (E) \equiv
\langle E \mid \eta \rangle = N {E}^{\alpha} exp ( - \beta E ) $,
then the conditions $\langle H \rangle = 1 $ and $\langle
{H}^{-1}\rangle < \infty $ lead to $ \beta - \frac{1}{2} = \alpha
> 0 $. This allow us to calculate all the matrix elements involving
time operator and even to study their classical limit as $ \hbar
\rightarrow 0 $ at the horizon.

In the limit $g\longrightarrow 0$, we have $\hat{H}
\longrightarrow H_0 = p^2/2$ and $\hat{t}\longrightarrow t_0$
where
\begin{equation}
\hat{t}_0 = - \frac{1}{2}(xp^{-1} + p^{-1}x)
\end{equation}

is exactly the time-of-arrival operator of Aharonov and Bohm
\cite{Aha}. The time operators $\hat t $ and $\hat t_{0} $ can
also be related to each other by means of a unitary operator \cite
{Mart} that transforms $ \hat H  \rightarrow \hat H_{0} $:

\beq
\hat {H}= U \hat {H_{0}} U^{\dagger}, \,\, \,\,\, \hat {t} =
U \hat{t_{0}} U^{\dagger}.
\eeq

The Aharonov-Bohn time operator $\hat{t}_0$ is not self-adjoint
and its eigenfunctions are not orthonormal.

\section{Conclusion}
In this paper, we have studied the  properties of a scalar field
in the near-horizon region of a massive Schwarzschild black hole.
The quantum Hamiltonian governing the near-horizon dynamics is
found to be  scale invariant and has the full conformal group as a
dynamical symmetry group. Using only the generators of the affine
group, we constructed the time operator near-horizon. The
self-adjointness of $\hat{H}$ and $\hat{t}$ is also discussed.

\section{Acknowledgment}
This work was supported by the Ministry of Science and Technology
of the Republic of Croatia under Contract No.0098004.

\end{document}